\documentclass[11pt]{article}
\usepackage{times}
\usepackage{epsfig}

\makeatletter

\DeclareSymbolFont{boldmath}{OML}{cmm}{b}{it}
\DeclareSymbolFontAlphabet{\mathb}{boldmath}
\DeclareMathAlphabet{\Bbb}{U}{msb}{m}{n}
\DeclareMathAlphabet{\euf}{U}{euf}{b}{n}
 
\DeclareMathSymbol{\balpha}{0}{boldmath}{"0B}
\DeclareMathSymbol{\bbeta}{0}{boldmath}{"0C}
\DeclareMathSymbol{\bgamma}{0}{boldmath}{"0D}
\DeclareMathSymbol{\bdelta}{0}{boldmath}{"0E}
\DeclareMathSymbol{\bepsilon}{0}{boldmath}{"0F}
\DeclareMathSymbol{\bzeta}{0}{boldmath}{"10}
\DeclareMathSymbol{\bfeta}{0}{boldmath}{"11}
\DeclareMathSymbol{\btheta}{0}{boldmath}{"12}
\DeclareMathSymbol{\biota}{0}{boldmath}{"13}
\DeclareMathSymbol{\bkappa}{0}{boldmath}{"14}
\DeclareMathSymbol{\blambda}{0}{boldmath}{"15}
\DeclareMathSymbol{\bmu}{0}{boldmath}{"16}
\DeclareMathSymbol{\bnu}{0}{boldmath}{"17}
\DeclareMathSymbol{\bxi}{0}{boldmath}{"18}
\DeclareMathSymbol{\bpi}{0}{boldmath}{"19}
\DeclareMathSymbol{\brho}{0}{boldmath}{"1A}
\DeclareMathSymbol{\bsigma}{0}{boldmath}{"1B}
\DeclareMathSymbol{\btau}{0}{boldmath}{"1C}
\DeclareMathSymbol{\bupsilon}{0}{boldmath}{"1D}
\DeclareMathSymbol{\bphi}{0}{boldmath}{"1E}
\DeclareMathSymbol{\bchi}{0}{boldmath}{"1F}
\DeclareMathSymbol{\bpsi}{0}{boldmath}{"20}
\DeclareMathSymbol{\bomega}{0}{boldmath}{"21}
\DeclareMathSymbol{\beps}{0}{boldmath}{"22}
\DeclareMathSymbol{\bthet}{0}{boldmath}{"23}
\DeclareMathSymbol{\bomeg}{0}{boldmath}{"24}
\DeclareMathSymbol{\bvphi}{0}{boldmath}{"27}

\@addtoreset{equation}{section}
\newcommand{\fref}[1]{figure~\ref{#1}} 
\newcommand{\eref}[1]{(\ref{#1})}
\newcommand{\ft}[2]{{\textstyle{{#1}\over{#2}}}}
\newcommand{\expo}[1]{e^{#1}} 
\newcommand{\del}{\partial} 
\newcommand{\dd}{\mathrm{d}}
 
\newcommand{\Tr}{\mathrm{Tr}} 
\newcommand{\Trr}[1]{\Tr(#1)}
\newcommand{\grpSL}{\mathsf{SL}}

\newcommand{\algsl}{\euf{sl}}

\newcommand{\RR}{\Bbb{R}} 
\newcommand{\N}{\mathcal{N}} 
\newcommand{\M}{\mathcal{M}} 
\newcommand{\U}{\mathcal{U}}
\newcommand{\Qsp}{{\mathcal{Q}}} 
\newcommand{\Psp}{{\mathcal{P}}}
\newcommand{\Rsp}{{\mathcal{R}}}
\newcommand{\QQsp}{{\widetilde\mathcal{Q}}} 
\newcommand{\PPsp}{{\widetilde\mathcal{P}}}
\newcommand{\RRsp}{{\widetilde\mathcal{R}}}
\newcommand{\sym}{\mathit{\Omega}}
\newcommand{\dif}{h}
\newcommand{\gam}{\bgamma} 
\newcommand{\one}{\mathbf{1}}
\newcommand{\ee}{\mathb{e}} 
\newcommand{\om}{\bomega}
\newcommand{\F}{\mathb{F}} 
\newcommand{\T}{\mathb{T}} 
\newcommand{\lie}{\mathcal{L}}
\newcommand{\fff}{\mathb{f}} 
\newcommand{\ggg}{\mathb{g}}
\newcommand{\llpar}{\mathb{k}}   
\newcommand{\llgen}{\blambda}    
\newcommand{\ltpar}{\mathb{n}}   
\newcommand{\ltgen}{\bzeta}      
\newcommand{\dfgen}{\xi}           
\newcommand{\uu}{\mathb{u}} 
\newcommand{\vv}{\mathb{v}} 
 
\newcommand{\xx}{\mathb{x}} 
\newcommand{\yy}{\mathb{y}}

\newcommand{\eps}{\varepsilon}
\newcommand{\newton}{G} 
\newcommand{\cut}{\lambda}

\makeatother

\begin{document}

\begin{flushright}
MZ-TH/99-04\\
gr-qc/9903040
\end{flushright}

\begin{center}
  \LARGE \textbf{On the relation between 2+1 Einstein gravity\\
    and Chern Simons theory}
\end{center}
 
\vspace*{8mm}

\begin{center}
  \textbf{Hans-J\"urgen Matschull}\\[2ex]
  Institut f\"ur Physik, Johannes Gutenberg-Universit\"at\\[1ex]
  Staudingerweg 7, 55099 Mainz, Germany\\[1ex]
  E-mail: matschul@thep.physik.uni.mainz.de
\end{center}

\vspace*{10mm}

\begin{abstract}
  A simple example is given to show that the gauge equivalence classes
  of physical states in Chern Simons theory are not in one-to-one
  correspondence with those of Einstein gravity in three spacetime
  dimensions. The two theories are therefore not equivalent.  It is
  shown that including singular metrics into general relativity has
  more, and in fact a quite counter-intuitive, impact on the theory than
  one naively expects.
\end{abstract}

\vspace*{10mm}

\section{Outline}
It is often argued that Einstein gravity on a given three dimensional
spacetime manifold $\M$ is equivalent to a Chern Simons theory on the
same manifold with the Poincar\'e group being the underlying gauge group
\cite{witten,carnel,matrev}. In fact, writing the first order Einstein
Hilbert action in the dreibein formulation of general relativity, one
finds that this is, up to a total derivative, equal to the Chern Simons
action. The dreibein and the spin connection are then interpreted as the
translational and rotational components of a Poincar\'e algebra valued
gauge field. The only difference between the two theories is that in
Einstein gravity the dreibein is restricted to be invertible, whereas no
such restriction exists in Chern Simons theory. In this sense, Chern
Simons theory can be thought of as an extension of Einstein gravity
including singular metrics, or vice versa Einstein gravity is considered
as a restricted version of Chern Simons theory.

The purpose of this article is to show that this equivalence holds at
the level of field configurations only. There is indeed a one-to-one
relation between off shell as well as on shell field configurations of
Einstein gravity, and those field configurations of Chern Simons theory
which have a non-vanishing dreibein determinant everywhere on $\M$.
However, there is no such relation at the level of \emph{gauge
  equivalence classes} of physical states.  After dividing out the gauge
symmetries of the two theories, one ends up with two different reduced
phase spaces. Both are finite dimensional manifolds, that is, both
theories are \emph{topological} in the sense that they have only
finitely many physical degrees of freedom.  But there is no one-to-one
correspondence between the reduced states of Chern Simons theory and
those of Einstein gravity.

The basic idea of the proof is the following. Assume that $\Qsp$ is the
configuration space of Chern Simons theory, that is, the set of all
Poincar\'e algebra valued one-forms on a given spacetime manifold $\M$.
They are typically subject to some boundary conditions, but these are
not of any importance here. Assume further, that there is an action
principle defined on $\Qsp$, which provides a set of field equations
that singles out a submanifold $\Psp\subset\Qsp$ of \emph{physical
  states}, or \emph{on shell} field configurations. This set of
solutions $\Psp$ can be identified with the \emph{physical phase space}
of Chern Simons theory.  The action defines a canonical symplectic
two-form $\sym_\Psp$ thereon.  It is highly degenerate, because only
finitely many dimensions of the infinite dimensional phase space $\Psp$
are actually physical degrees of freedom. All other dimensions, namely
the null directions of $\sym_\Psp$, are directions of gauge
transformations.

In general, two states $\phi_1,\phi_2\in\Psp$ are gauge equivalent,
denoted by $\phi_1\sim\phi_2$, if they lie on the same \emph{null orbit}
of $\sym_\Psp$, that is, if they can be joint by a curve in $\Psp$ whose
tangent vector lies in the kernel of the two-form $\sym_\Psp$ everywhere
along the curve. This is the definition of a \emph{smoothly generated}
gauge symmetry. In the case of Chern Simons theory, the gauge symmetries
are the local Poincar\'e transformations.  Depending on the topology of
the spacetime manifold $\M$, some of them might not be smoothly
generated. These are called \emph{large} local Poincar\'e
transformations. Large gauge transformations are symmetries of the
physical phase space that map those states onto each other, which are
physically indistinguishable but do not lie on the same null orbit of
the symplectic two-form. It is thereby important to note that the
existence of large gauge symmetries can \emph{not} be inferred from the
given action.

Instead, it depends on the physical interpretation of the model whether
two states related by a large transformation should be considered as
gauge equivalent or not. In particular, one has to decide which
quantities can be measured and which cannot. Depending on this choice,
an equivalence class of physical states, in the following also called a
\emph{gauge orbit}, consist of one or several null orbits of
$\sym_\Psp$. If there are no large gauge symmetries, then each gauge
orbit is a connected submanifold of $\Psp$ and consists of exactly one
null orbit of $\sym_\Psp$.  Otherwise, the gauge orbits are disconnected
submanifolds of $\Psp$ consisting of more than one null orbit of
$\sym_\Psp$. In any case, the reduced phase space $\Rsp=\Psp/{\sim}$ is
defined to be the quotient space of the physical phase space modulo
gauge symmetries. Provided that the gauge orbits on $\Psp$ behave
regularly enough, it is a manifold whose dimension is equal to the rank
of $\sym_\Psp$. A unique, non-degenerate symplectic two-form $\sym_\Rsp$
can then be defined on $\Rsp$, such that $\sym_\Psp$ is its pullback on
$\Psp$.

By definition, each gauge invariant function on $\Psp$, usually called
an \emph{observable}, can be written as a function on $\Rsp$ and vice
verse. Hence, the reduced phase space $\Rsp$ contains the full
physically relevant information about the system. In other words, if
Einstein gravity is equivalent to Chern Simons theory, or a restricted
version thereof, then the reduced phase space of Einstein gravity should
be equal to, or a subset of the reduced phase space of Chern Simons
theory.  But this, as we are now going to see, is not the case. Let me
briefly explain what goes wrong, before making things more explicit. In
the configuration space $\Qsp$ of Chern Simons theory, there is a
\emph{submanifold of singular metrics}, consisting of all field
configurations with vanishing dreibein determinant at some point in
$\M$. The configuration space $\QQsp$ of Einstein gravity is obtained
from $\Qsp$ but taking away this submanifold.

\begin{figure}[t]
  \begin{center}
    \epsfbox{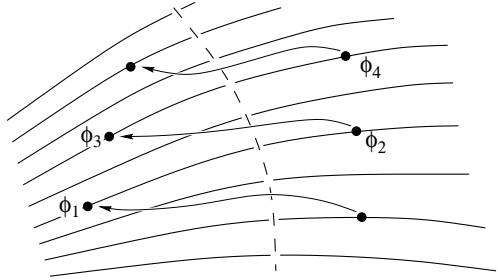}
    \caption{The gauge orbits (solid lines) and the submanifold of
      singular metrics (dashed line) in the physical phase space $\Psp$
      of Chern Simons theory. The states $\phi_1$ and $\phi_2$ are
      related by a local Poincar\'e transformation in Chern Simons
      theory, but they are not gauge equivalent in Einstein gravity.
      Instead, $\phi_2$ is related to $\phi_3$ by a large
      diffeomorphism.}
    \label{orb}
  \end{center}
  \hrule
\end{figure}

The action and also the equations of motion remain the same, which
implies that the same holds for the physical phase space $\PPsp$ of
Einstein gravity. It is the subset of $\Psp$ obtained by taking away the
submanifold of singular metrics. But now something crucial is happening
to the gauge orbits. First of all, we have $\sym_\Psp=\sym_\PPsp$, and
therefore the null orbits of $\sym_\PPsp$ in Einstein gravity are the
same as the null orbits of $\sym_\Psp$ in Chern Simons theory. The
problem is that they intersect with the submanifold of singular metrics.
This is illustrated in \fref{orb}. As a consequence, when the
submanifold is taken away, the null orbits fall apart into disconnected
components. For example, the states $\phi_1$ and $\phi_2$, which are
related by a smoothly generated gauge symmetry in Chern Simons theory,
no longer lie on the same null orbit of $\sym_\PPsp$, and thus they are
not related by a smoothly generated gauge symmetry in Einstein gravity.
As a result, the reduced phase space $\RRsp=\PPsp/{\sim}$ of Einstein
gravity is different from that of Chern Simons theory,
$\Rsp=\Psp/{\sim}$.

Now, one might argue that the equivalence of the two theories can be
easily restored by considering the states $\phi_1$ and $\phi_2$ in
Einstein gravity as being related by a \emph{large} gauge symmetry. Each
gauge orbit then consists of several disconnected parts, namely exactly
those that were formerly, in Chern Simons theory, a single null orbit.
In Einstein gravity, we would then say that $\phi_1$ and $\phi_2$, or
$\phi_3$ and $\phi_4$, are related by a large local Poincar\'e
transformation. From the mathematical point of view, this would be
perfectly consistent, and the resulting reduced phase spaces
$\Rsp=\Psp/{\sim}$ and $\RRsp=\PPsp/{\sim}$ would indeed be identical.
The only difference would be that a reduced state of Einstein gravity
has less representatives in $\PPsp$ than the corresponding reduced state
in Chern Simons theory has in $\Psp$. However, remember that the
question whether large gauge symmetries exist was a physical question
rather than a mathematical one. So, what we should ask is whether the
states $\phi_1$ and $\phi_2$ are physically indistinguishable.

Unfortunately, this is not the case. A simply example will show that
declaring two such states as gauge equivalent in Einstein gravity leads
to a straight contradiction to our intuitive understanding of the very
basic principles of general relativity. To express it more drastically,
if nobody would know about the relation between Einstein gravity and
Chern Simons theory, then nobody would have ever suggested to declare
the states $\phi_1$ and $\phi_2$ in \fref{orb} to be gauge equivalent.
Or, phrased differently, given only the phase space $\PPsp$ of Einstein
gravity, why should we glue two \emph{a priori} unrelated regions of
this space together along the dashed line, in the way shown in the
figure, especially if we do not know how to interpret the states
\emph{on} that line? In any case, we have to conclude that the
similarity between Einstein gravity and Chern Simons theory can be quite
useful, but when we speak about gauge symmetries and the reduced phase
space, we have to decide which of the two theories we are dealing with.

Without making this decision, some authors run into yet another problem
\cite{baismul}. It arises if one tries to include too many features of
both Einstein gravity and Chern Simons theory into a kind of mixture of
the two theories, which then has too many large gauge symmetries. In
Einstein gravity, an essential part of the gauge group are the
diffeomorphisms of the underlying spacetime manifold $\M$. In the spirit
of \emph{general relativity}, one assumes that the individual points of
$\M$ do not have any particular physical meaning. This implies that
every transformation that interchanges them without destroying the
differentiable structure of $\M$ should be considered as a gauge
symmetry. For a generic manifold $\M$, there are smoothly generated
diffeomorphisms as well as large ones. Now, the orbits of the smoothly
generated diffeomorphisms in phase space are included in the orbits of
smoothly generated local Poincar\'e transformations. This is because the
infinitesimal generator of any diffeomorphisms can be written as a local
Poincar\'e transformation with a special parameter.

However, it turns out that the large diffeomorphisms of $\M$ have
nothing to do with the local Poincar\'e transformations. They act on the
physical phase space in a completely different way. For example, the
state $\phi_2$ in the figure could be mapped onto $\phi_3$ by some large
diffeomorphism, lying on a different null orbit of $\sym_\Psp$. In
Einstein gravity, including the large diffeomorphism as gauge symmetries
does not course a problem. What happens is that several null orbits of
$\sym_\PPsp$ are united into a single gauge orbit. However, the null
orbits to the left and to the right of the dashed line in the figure are
thereby identified in a different way, as compared to the way they are
glued together by including the submanifold of singular metrics. A
problem arises if we now consider the large diffeomorphisms as gauge
symmetries in Chern Simons theory, or if \emph{both} the large
Poincar\'e transformations \emph{and} the large diffeomorphisms are
considered as gauge symmetries in Einstein gravity.

In both cases, the states $\phi_1$ through $\phi_4$ in the figure all
become gauge equivalent, and we can extend this sequence into both
directions, as indicated by the arrows. Now, assume that the plane shown
in the figure is actually a part of a cylinder, and after a finite
number of steps we get back to this region, but we never again hit the
null orbit we started from. The gauge orbit then fills the plane
densely, and the resulting reduced phase space is no longer a manifold.
It is in particular not Hausdorff, because two such gauge orbits which
both fill the plane densely cannot be separated by open neighbourhoods.
Using the example already mentioned above, we shall see that this is a
quite typical situation. It arises because one tries to keep too many
features of both theories. If one either excludes singular metrics,
sticking to Einstein gravity, or does not regard large diffeomorphisms
as gauge symmetries, for which there is actually no motivation in Chern
Simons theory, then this problem is absent.

\section{Chern Simons theory and Einstein gravity}
Let us now make things more explicit. In the following, we assume that
$\M$ is an orientable, three dimensional manifold with a trivial tangent
bundle, equipped with a Levi Civita tensor $\eps^{\mu\nu\rho}$.  On
$\M$, we introduce two $\algsl(2)$ valued one-forms $\om_\mu$ and
$\ee_\mu$. In Chern Simons theory, these are interpreted as the
rotational and the translational components of a Poincar\'e algebra
valued gauge field. Note that $\algsl(2)$ is the spinor representation
of the three dimensional Lorentz algebra, but it is also, as a vector
space, isomorphic to Minkowski space. The Poincar\'e field strength also
splits into an $\algsl(2)$ valued rotational component
\begin{equation}
  \label{curv}
  \F_{\mu\nu} = \del_\mu \om_\nu - \del_\nu \om_\mu 
                + [\om_\mu,\om_\nu],
\end{equation}
and an also $\algsl(2)$ valued translational component
\begin{equation}
  \label{tors}
  \T_{\mu\nu} = \del_\mu \ee_\nu - \del_\nu \ee_\mu 
               + [\om_\mu,\ee_\nu] - [\om_\nu,\ee_\mu] . 
\end{equation}
In Einstein gravity, $\om_\mu$ is interpreted as the spin connection and 
$\ee_\mu$ is the dreibein, defining the metric and the determinant, 
\begin{equation}
  \label{metric}
  g_{\mu\nu} = \ft12\Trr{\ee_\mu\ee_\nu}, \qquad
  e = \ft1{12} \eps^{\mu\nu\rho} \, \Trr{\ee_\mu\ee_\nu\ee_\rho}.
\end{equation}
The latter is required to be different from zero everywhere on $\M$.
Hence, the configuration space $\Qsp$ of Chern Simons theory is the set
of all field configurations $(\om_\mu,\ee_\mu)$ on $\M$, and the
configuration space $\QQsp$ of Einstein gravity is the set of all
$(\om_\mu,\ee_\mu)$ with $e\neq0$. In both cases, they are subject to
some additional restrictions such as smoothness and fall off conditions
at infinity. A more detailed discussion, taking into account all such
boundary conditions for a specific model, will be given in
\cite{matmul}. Here, we only need to know that all these extra conditions
can be chosen to be the same in both theories, such that $\QQsp$ is the
subset of $\Qsp$ with the singular metrics excluded. The vacuum Einstein
Hilbert action can then be written as
\begin{equation}
  \label{eh}
  \frac1{16\pi\newton} \int \dd^3 x \, \eps^{\mu\nu\rho} \,
                  \Trr{\ee_\mu\F_{\nu\rho}}.
\end{equation}
Up to a total derivative, this coincides with the Chern Simons action
\cite{witten,carnel}. The equations of motions are
\begin{equation}
  \label{eom}
  \F_{\mu\nu} = 0, \qquad
  \T_{\mu\nu} = 0,
\end{equation}
defining the physical phase space $\Psp$, respectively $\PPsp$. In Chern
Simons theory, they state that the Poincar\'e field strength vanishes.
In Einstein gravity they are interpreted as the \emph{torsion} the
\emph{curvature} equation, stating that the dreibein is covariantly
constant and that the metric is flat. All we need to know about the
symplectic form $\sym_\Psp=\sym_\PPsp$ is that its null directions are
the generators of the local symmetries of the action \eref{eh}.  These
are the \emph{local Poincar\'e transformations},
\begin{equation}
  \label{pnc}
  \delta\om_\mu = \del_\mu \llgen + [\om_\mu,\llgen],\qquad
  \delta\ee_\mu = \del_\mu \ltgen + [\om_\mu,\ltgen] + [\ee_\mu,\llgen],
\end{equation}
where $\llgen:\M\to\algsl(2)$ is the generating parameter of a
\emph{local Lorentz rotation} and $\ltgen:\M\to\algsl(2)$ that of a
\emph{local translation}. It is easy to see that the generators of
spacetime diffeomorphisms are included in these transformations. If we
choose the parameters to be
\begin{equation}
  \label{par}
  \llgen=\dfgen^\mu\om_\mu, \qquad
  \ltgen=\dfgen^\mu\ee_\mu,
\end{equation}
for some vector field $\dfgen^\mu$ on $\M$, then the transformation
\eref{pnc} becomes
\begin{equation}
  \label{lie}
  \delta\om_\mu = \lie_\dfgen \om_\mu 
                + \F_{\mu\nu} \, \dfgen^\nu ,\qquad
  \delta\ee_\mu = \lie_\dfgen \ee_\mu 
                + \T_{\mu\nu} \, \dfgen^\nu ,
\end{equation}
where $\lie_\dfgen$ denotes the Lie derivative with respect to the
vector field $\dfgen^\mu$. As the field strength, respectively the
curvature and the torsion vanishes on the physical phase space, this is
equal to the infinitesimal generator of a diffeomorphism of $\M$ with
generating vector field $\dfgen^\mu$. What we also need in the following
are the finite versions of the gauge transformations. A finite local
Lorentz rotation is given by
\begin{equation}
  \label{lor}
  \om_\mu \mapsto \llpar^{-1} (\del_\mu + \om_\mu ) \llpar, \qquad
  \ee_\mu \mapsto \llpar^{-1} \ee_\mu \llpar,
\end{equation}
where $\llpar:\M\to\grpSL(2)$ is a parameter field taking values in the
spinor representation of the Lorentz group $\grpSL(2)$. At this point,
however, we already have to be a little bit more careful. Not every
transformation of the form \eref{lor} can be written as a smooth
deformation of the fields generated by \eref{pnc}. This is the case if
and only if the map $\llpar:\M\to\grpSL(2)$ is \emph{contractible}, that
is, if it can be smoothly deformed into the trivial map $\llpar=\one$.
Otherwise the given transformation is a \emph{large} local Lorentz
rotation. So, already here we have to make a decision, namely whether to
consider large local Lorentz rotations as gauge symmetries or not.

Both in Einstein gravity as well as in Chern Simons theory, it is
physically reasonable to include them. Doing this essentially means that
the rotational component of the gauge field, respectively the spin
connection, can only be measured by parallel transport of vectors or
spinors along closed loops in $\M$. One can show that this leads to the
conclusion that all transformations of the form \eref{lor} are gauge
symmetries \cite{matwel}. But this is not the crucial point regarding
the difference between Einstein gravity and Chern Simons theory, and
therefore it is actually not relevant how to deal with the large local
Lorentz rotations. The more interesting gauge symmetries are the local
translations. A finite local translation is given by
\begin{equation}
  \label{trn}
  \om_\mu \mapsto \om_\mu ,\qquad
  \ee_\mu \mapsto \ee_\mu + \del_\mu \ltpar + [\om_\mu,\ltpar],
\end{equation}
where $\ltpar:\M\to\algsl(2)$ is actually the same parameter field as
$\ltgen$ in \eref{pnc}, but now it is a finite parameter rather than an
infinitesimal generator. One can easily check that together with the
local Lorentz rotations \eref{lor}, the local translation form a local
Poincar\'e group. It is also useful to note that there are no
\emph{large} local translations in Chern Simons theory. Every
transformation of the form \eref{trn} can be written as a smooth
deformation by simply replacing $\ltpar$ by $\epsilon\,\ltpar$, with
$0\le\epsilon\le1$. For Einstein gravity, things are a bit more evolved,
as we shall see below.  Before coming to this, let us also write down
the action of a diffeomorphism $\dif:\M\to\M$ on the field
configuration. The one-forms are thereby replaced by their pullbacks, so
that the finite version of \eref{lie} reads
\begin{equation}
  \label{dif}
  \om_\mu \mapsto \dif_*\om_\mu , \qquad
  \ee_\mu \mapsto \dif_*\ee_\mu .
\end{equation}
From the relations between the infinitesimal generators of local
Poincar\'e transformations \eref{pnc} and diffeomorphisms \eref{lie}, we
know that the action of every \emph{smoothly generated} diffeomorphism
\eref{dif} can be written as a combination of a local Lorentz rotation
\eref{lor} and a local translation \eref{trn}, both being smoothly
generated. This is valid in Einstein gravity as well as in Chern Simons
theory. In Einstein gravity, however, there is yet another relation
between the infinitesimal generators which is not present in Chern
Simons theory. If the dreibein is invertible, then \eref{par} provides a
one-to-one relation between the generating parameter $\ltgen$ of local
translations and the generating vector field $\dfgen^\mu$ of spacetime
diffeomorphisms. Hence, in Einstein gravity we can consider the
spacetime diffeomorphisms instead of the local translations as the
basic gauge symmetries.

Regarding the smoothly generated ones, this is just a reparametrization.
However, it leads us to a different set of large transformations. To see
that the situation is indeed as shown in \fref{orb}, consider the action
of a local translation \eref{trn} in Einstein gravity. Given a field
configuration and a parameter $\ltpar$, there is no guarantee that the
transformed dreibein is still invertible. In fact, it is easy to show
that for any given field configuration it is also always possible to find
a parameter $\ltpar$ such that the transformed dreibein becomes
singular. So, the gauge orbits indeed intersect with the submanifold of
singular metrics, as indicated in the figure. Moreover, given two states
$\phi_1$ and $\phi_2$ with non-singular metric, and related by a local
translation \eref{trn} for some parameter $\ltpar$, the simple trick to
show that they can be smoothly transformed into each other does not
work, because when we replace $\ltpar$ by $\epsilon\,\ltpar$, then some
of the intermediate states might have a singular metric.

This does not yet prove that the gauge orbits fall apart into
disconnected components, because the phase space is infinite
dimensional, and therefore the two dimensional picture does not
necessarily give the correct impression. There might still be a null
curve of $\sym_\PPsp$ that connects $\phi_1$ and $\phi_2$ without
passing through the submanifold of singular metrics. A possible way to
prove that such a path does not exist is to show that the states
$\phi_1$ and $\phi_2$ are not related by a combination of a
diffeomorphism and a local Lorentz rotation. They are then, in
particular, not related by a smoothly generated diffeomorphism and a
local Lorentz rotation. On the other hand, we know that in Einstein
gravity every smoothly generated gauge symmetry can be written as a
combination of a smoothly generated diffeomorphism and a local Lorentz
rotations. If this is not the case for the two given states, it follows
that they are not related by a smoothly generated gauge symmetry, and
therefore they do not lie on a connected component of a gauge orbit.

From this we have to conclude that in Einstein gravity some of the local
translations are \emph{large}, and we can ask the question whether these
should be considered as gauge symmetries or not. It is important to note
that in Chern Simons theory this question never arises, because all
local translations are smoothly generated. Hence, if we decide
\emph{not} to consider the large local translation as gauge symmetries
of Einstein gravity, and we shall shortly see that our intuitive
understanding of general relativity forces us to do so, then the two
theories are not equivalent. We shall give an explicit example of two
states $\phi_1$ and $\phi_2$, related by a local translation, but
representing two spacetimes which are obviously not diffeomorphic and
therefore physically clearly distinguishable. From this we infer that,
first of all, the two states are not related by a smoothly
generated gauge symmetry, and secondly, that we must not consider large
translations as gauge symmetries of Einstein gravity if we want to
stick to its original physical interpretation.

\section{The counter example} 
\begin{figure}[t]
  \begin{center}
    \epsfbox{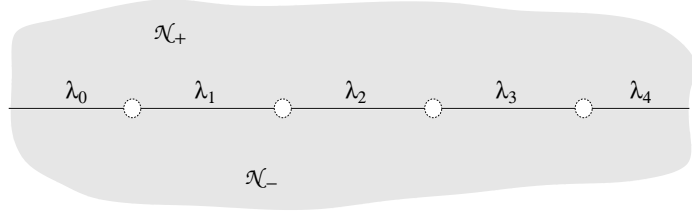}
    \caption{The space manifold $\N$ is a plane with four punctures. It
      can be covered by two simply connected regions $\N_\pm$
      overlapping along their common boundary, which consists of five
      lines $\cut_0$ through $\cut_4$.}
    \label{mfd}
  \end{center}
  \hrule
\end{figure}

To give the explicit example, let us first have a closer look at the
solutions to the field equations \eref{eom}. It is well known that
locally, within a simply connected region $\U\subset\M$, they can be
parametrized by two fields $\ggg:\U\to\grpSL(2)$ and
$\fff:\U\to\algsl(2)$, such that \cite{witten,matrev,grinar}
\begin{equation}
  \label{sol}
  \om_\mu = \ggg^{-1} \del_\mu \ggg, \qquad
  \ee_\mu = \ggg^{-1} \del_\mu \fff \, \ggg.
\end{equation}
In Chern Simons theory, this means that locally the fields are
\emph{pure gauge}. In fact, setting $\llpar=\ggg^{-1}$ in \eref{lor} and
then $\ltpar=-\fff$ in \eref{trn}, we can always transform the fields
into the trivial solution $\om_\mu\mapsto0$ and $\ee_\mu\mapsto0$. It
follows, in particular, that there are no local physical degrees of
freedom. In Einstein gravity, there is a restriction on $\fff$, and the
interpretation of the parameters is as follows. The spacetime metric and
the dreibein determinant can be written as
\begin{equation}
  \label{emb}
  g_{\mu\nu} = \ft12\Trr{\del_\mu\fff\,\del_\nu\fff}, \qquad
  e = \ft1{12} \eps^{\mu\nu\rho} \,
                \Trr{\del_\mu\fff\,\del_\nu\fff\,\del_\rho\fff}.
\end{equation}
The second equation states that the dreibein determinant is equal to the
Jacobian of $\fff$. If this is non-zero, then $\fff$ is locally
one-to-one. Hence, in contrast to Chern Simons theory, $\fff$ is not
completely arbitrary. According to the first equation, it then provides
an \emph{isometric embedding} of the simply connected subset $\U$ of
spacetime into Minkowski space. The components of the vector $\fff$ are
called \emph{Minkowski coordinates} on $\U$. So, even though we cannot
transform to $\om_\mu\mapsto0$ and $\ee_\mu\mapsto0$, because then the
metric is singular, we can still see that there are no local
gravitational degrees of freedom, because the spacetime is always flat.
The action of the gauge symmetries on the fields $\ggg$ and $\fff$ is as
follows.  Under a local Lorentz rotation with parameter
$\llpar:\M\to\grpSL(2)$, they transform as
\begin{equation}
  \label{emb-lor}
  \ggg \mapsto \ggg \, \llpar, \qquad
  \fff \mapsto \fff.
\end{equation}
For a local translation with parameter $\ltpar:\M\to\algsl(2)$, we find
that
\begin{equation}
  \label{emb-trn}
  \ggg \mapsto \ggg, \qquad
  \fff \mapsto \fff + \ggg \, \ltpar \, \ggg^{-1}.
\end{equation}
And finally, a spacetime diffeomorphism $\dif:\M\to\M$ acts as
\begin{equation}
  \label{emb-dif}
  \ggg \mapsto \ggg \circ \dif, \qquad
  \fff \mapsto \fff \circ \dif.
\end{equation}
To specify a physical state completely, using the field $\ggg$ and $\fff$
as parameters, we first have to fix a spacetime manifold $\M$, cover it
by a set of simply connected regions, and define the fields $\ggg$ and
$\fff$ independently within in these regions. Let us do this for a
special example. As a spacetime manifold, we choose the direct product
$\M=\RR\times\N$ of a real line, which is going to represent the time
coordinate $t$, with the space manifold $\N$ shown in \fref{mfd}. It is
a plane with four punctures, that is, four points are taken away. It is
not simply connected, but it can be covered by two simply connected,
closed subsets $\N_\pm$, overlapping along a sequence of five lines
$\cut_k$, with $k=0,\dots,4$, as indicated in the figure.

A physical state can then be parametrized by two pairs of fields
$\ggg_\pm:\M_\pm\to\grpSL(2)$ and $\fff_\pm:\M_\pm\to\algsl(2)$, where
$\M_\pm=\RR\times\N_\pm$. These fields are subject to a boundary
condition on the overlap surfaces $\RR\times\cut_k$ of $\M_+$ and
$\M_-$. For the dreibein and the spin connection to be well defined, the
right hand sides of \eref{sol} must be the same in $\M_+$ and $\M_-$.
This is the case if and only if the values of $\ggg_\pm$ and $\fff_\pm$
in the overlap region are related by
\begin{equation}
  \label{cut}
  \ggg_- = \uu_k\!\!^{-1} \ggg_+ , \qquad
  \fff_- = \uu_k\!\!^{-1} ( \fff_+ - \vv_k ) \, \uu_k,
\end{equation}
for some constants $\uu_k\in\grpSL(2)$ and $\vv_k\in\algsl(2)$. The
second equation states that $\fff_-$ is related to $\fff_+$ by a rigid
Poincar\'e transformation, that is, an isometry of the embedding
Minkowski space, consisting of a translation $\vv_k$ and a Lorentz
rotation $\uu_k$. This implies that the spacetime metric induced by
$\fff_+$ is the same as that induced by $\fff_-$. The first equation
then ensures that not only the metric but also the dreibein and the spin
connection are continuous on $\M$. If the overlap region splits into 
several disconnected components, then there are independent constants
for each part. In our case, we have five independent \emph{transition
  functions} $\uu_k$ and $\vv_k$, with $k=0,\dots,4$.

Using this, we can formulate the following statement, which holds in
Chern Simons theory, but not in Einstein gravity. Given two physical
states, parametrized by fields $\ggg$ and $\fff$, respectively $\ggg'$
and $\fff'$, such that the transition functions $\uu_k$ and $\vv_k$
coincide, then the two states are gauge equivalent. The proof is quite
simple.  Define a field $\llpar=\ggg^{-1}\ggg'$. By assumption, this is
a continuous field on $\M$, which follows from the first equation in
\eref{cut}. Inserted into \eref{emb-lor}, it transforms $\ggg$ into
$\ggg'$ by a local Lorentz rotation. Next, assume that $\ggg=\ggg'$ and
define $\ltpar=\ggg^{-1}(\fff'-\fff)\ggg$.  According to the second
equation in \eref{cut}, this is again a continuous field on $\M$, and
when we take this as a parameter of a local translation \eref{emb-trn},
$\fff$ is transformed into $\fff'$.  Hence, given two states with
coinciding transition functions, we can always find a combination of a
local Lorentz rotation and a local translation that maps the two states
onto each other.

What remains to be done to prove the non-equivalence of Einstein gravity
and Chern Simons theory is to find two states of Einstein gravity with
coinciding transition functions, which are however physically
distinguishable and therefore not gauge equivalent. They are then
related by a local translation. But this cannot be a smoothly generated
one because then the two states would also be related by a
diffeomorphism. Hence, it must be a large local translation, and we have
to conclude that the large local translations are not to be considered
as gauge symmetries of Einstein gravity. To give two such states
explicitly, we have to introduce some more notation. Let $\gam_a$, with
$a=0,1,2$, be some orthonormal basis of $\algsl(2)$, such that
$\gam_a\gam_b=\eta_{ab}\one-\eps_{abc}\gam^c$, where $\eta_{ab}$ is the
Minkowski metric with signature $(-,+,+)$ and $\eps_{abc}$ is the Levi
Civita tensor with $\eps_{012}=-1$. This implies, for example, that
conjugation of a vector $\vv$ with a group element
$\expo{\alpha\gam_0}\in\grpSL(2)$ provides a spatial rotation of $\vv$
by $2\alpha$. This is all we need to know about the algebra of the gamma
matrices. To define an explicit solution to the field equations, we
first specify the following transition functions,
\begin{equation}
  \label{uu-k}
  \uu_0 = \expo{-2m\gam_0},\quad
  \uu_1 = \expo{- m\gam_0},\quad
  \uu_2 = \one,\quad
  \uu_3 = \expo{ m\gam_0},\quad
  \uu_4 = \expo{2m\gam_0},
\end{equation}
where $m=\pi/12$. These group elements represent spatial rotations by
$-60^\circ$, $-30^\circ$, $0^\circ$, $30^\circ$, and $60^\circ$. Then,
we choose two arbitrary, smooth fields $\ggg_\pm:\M_\pm\to\grpSL(2)$,
such that the boundary conditions \eref{cut} are satisfied. Their
precise form does not matter. They are all related by local Lorentz
rotations anyway. All we need to know is that such fields exist, which
can be easily verified. Finally, we have to define the fields
$\fff_\pm$. We decompose them into time and space components,
\begin{equation}
  \label{fex}
  \fff_\pm(t,z) = t \, \gam_0 + \vec f_\pm(z) \cdot \vec\gam,
\end{equation}
where $\vec f_\pm(z)=(f_1(z),f_2(z))$ maps the points $z\in\N_\pm$ onto
points in the Euclidean plane, considered as the spatial plane in
Minkowski space, which is spanned by $\vec\gam=(\gam_1,\gam_2)$. If the
functions $\vec f_\pm$ are independent of $t$, it follows from
\eref{fex} and \eref{emb} that the resulting spacetime is static, and
that the time coordinate $t$ is the physical time. It is then sufficient
to specify the spatial metric to fix the geometry of the whole spacetime
manifold.
\begin{figure}[t]
  \begin{center}
    \epsfbox{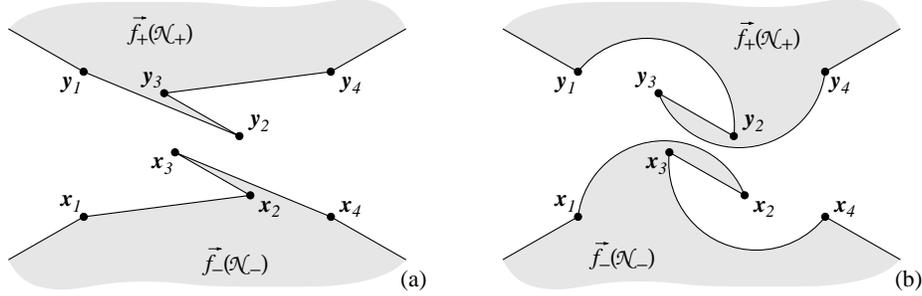}
    \caption{The images of the maps $\vec f_\pm$ on the spatial plane in 
      Minkowski space. The geometry of the space manifold can be read
      off by gluing the two flat surfaces together along their edges.
      The space obtained in (a) is a cone with four tips in a row, which
      is different from that in (b), where the tips form a
      parallelogram.}
    \label{img}
  \end{center}
  \hrule
\end{figure}

We choose the maps $\vec f_\pm$ to be one-to-one, and such that their
images $\vec f_\pm(\N_\pm)$ are the subsets of the spatial plane shown
in \fref{img}(a). It is, once again, not important to know what these
maps precisely looks like. In fact, the images determine them up to a
diffeomorphism of $\N$, which can even be shown to be a smoothly
generated one. Hence, we have the statement that given the transition
functions $\uu_k$ and the images of $\vec f_\pm$, the state is fixed up
to a local Lorentz rotation and a smoothly generated diffeomorphisms.
What is not immediately obvious is that, given the images, we can choose
$\vec f_\pm$ such that the boundary conditions \eref{cut} are satisfied.
The condition for this to be possible is that the edges of the upper and
lower surface in the figure are mapped onto each other by isometries of
the Euclidean plane, consisting of a translation and a rotation by
$-60^\circ$, $-30^\circ$, $0^\circ$, $30^\circ$, and $60^\circ$,
respectively. In particular, we must have the following relations
between the corner points $\xx_k$ and $\yy_k$ in the figure and the
transition functions,
\begin{equation}
  \label{map}
  \xx_k = \uu_k\!\!^{-1} ( \yy_k - \vv_k ) \, \uu_k, \qquad
  \xx_k = \uu_{k-1}\!\!\!\!^{-1} 
            ( \yy_k - \vv_{k-1} ) \, \uu_{k-1},
\end{equation}
for $k=1,\dots,,4$. These relations are indeed satisfied, with the
$\uu_k$ taken from \eref{uu-k} and for some suitable chosen values of
$\vv_k$. The first equation follows from evaluating \eref{cut} at the
beginning of the line $\cut_k$, and the second from the same condition
at the end of $\cut_{k-1}$. 

To see what the geometry of the spacetime manifold looks like, we have
to cut out the surface in \fref{img}(a), and glue them together along
their edges. Note that they always fit together, because the
corresponding edges of the two half spaces are mapped onto each other by
isometries of the embedding Minkowski space. In particular, we always
obtain a locally flat spacetime. The resulting space manifold can be
described as a cone with four tips, arranged in a row.  Its total
deficit angle is $8m=120^\circ$, and at each tip there is a deficit
angle of $2m=30^\circ$. The spacetime obtained by taking the direct
product with a real line can be interpreted as a universe containing
four point particles with mass $m$, being at rest with respect to each
other. The mass is thereby measured in units of the Planck mass
$1/(4\pi\newton)$ \cite{matwel,thooft,matmul}.

It is now possible to construct a \emph{different} universe, containing
the same particles, which is not diffeomorphic to the first, but with
all transition function being the same, so that the two states of
Einstein gravity are related by a large translation. It is thereby
sufficient to note that the transition functions are already determined
by the corner points $\xx_k$ and $\yy_k$ in the figure, and the
directions of the edges extending to infinity. This follows from
\eref{map}, or simply from the fact that a Poincar\'e transformation,
which is in this case just an isometry of the Euclidean plane, is fixed
if its action on two points is known. If we, for example, choose the
maps $\vec f_\pm$ such that their images are the surfaces in
\fref{img}(b), then the transition functions remain the same, because
the corner points and the directions of the edges extending to infinity
are unchanged.

In fact, one can easily see that the rotations and translations needed
to map the corresponding edges of the upper surface onto those of the
lower surface as the same in \fref{img}(a) and (b). But if we now cut
out the surfaces and glue them together along the edges, then we find
that the particles are no longer arranged in a row. Instead, they form a
parallelogram. It is immediately obvious that the two static spacetimes
are not diffeomorphic, and therefore they are not gauge equivalent
states of Einstein gravity. This completes the proof that there are
states on the gauge orbits of Chern Simons theory $\phi_1$ and $\phi_2$,
as shown in \fref{orb}, which do not lie on the same gauge orbit of
Einstein gravity, and it also explains why it is contradictory to the
usual interpretation of general relativity to consider large
translations as gauge symmetries of Einstein gravity.

An important point in this consideration is that it does not depend on
the boundary conditions imposed on the fields. They usually play a very
important role in the precise definition of the phase space, because
they have to guarantee that the action is finite, and they also have
some impact on what is a gauge symmetry and what is not \cite{matmul}.
For example, a transformation with a \emph{constant} parameter $\llgen$
or $\ltgen$ is typically not a gauge symmetry when certain fall off
conditions are imposed on the fields.  However, all this has nothing to
do with our simple counter example. The equivalence of the two theories
can not be restored by imposing boundary condition on the fields,
neither at infinity nor in the neighbourhood of the punctures. The
crucial transformation, namely the large translation that takes us from
(a) to (b) in \fref{img}, takes place within a \emph{compact} region of
$\N$. The fields $\vec f_\pm$ can be chosen to be the same in (a) and
(b), say, outside the finite region shown in the figure, and by slightly
deforming the images in the neighbourhoods of the corner points $\xx_k$
and $\yy_k$, we can even achieve that the fields are unchanged in a
finite neighbourhood of the particles.

What remains to be seen is that regarding the large diffeomorphisms as
gauge symmetries of Chern Simons theory, or to consider both large
diffeomorphisms and large translations as gauge symmetries of Einstein
gravity, makes the gauge orbits behave such that certain regions of the
phase space are filled densely by a single orbit, so that the reduced
phase space becomes non-Hausdorff. To see this, observe that there are
infinitely many inequivalent ways to draw the edges of the surfaces in
\fref{img}. The only conditions are that the corners are connected in
the correct ordering, and that the corresponding edges of the lower
surface are mapped onto those of the upper surfaces by the given
isometries \eref{cut}. But even if we, for example, fix the edges
extending to infinity and the one between the second and the third
corner, then there is still an infinite series of possibilities, two of
which are shown. The edge starting at $\xx_1$, for example, can wind
around $\xx_3$ any number of times before it gets to $\xx_2$. We can
label the different states by this winding number. All these states are
then related to each other by local translations.

How do the resulting spacetimes look like? They always consist of four
identical point particles being at rest. Two of them, number $1$ and
$4$, are at fixed positions, and the other two, number $2$ and $3$, are
located in between them, with varying angular orientation. Each time we
increase the winding number, the inner particles are rotated by $4m$
with respect to the outer ones. For the example given, we have
$4m=60^\circ$, which means that increasing the winding number by six, or
three if we consider the particles as being identical, leads us back to
the same geometry. A closer analysis shows that the two states are then
related by a large diffeomorphism. But what happens if the mass $m$ of
the particles is an irrational multiple of $\pi$? We can then make the
same construction, but the possible angular orientations take countably
many different values for the different winding numbers. The possible
locations of the inner particles with respect to the outer ones fill a
circle densely.

In other words, a single gauge orbit fills a certain region of the phase
space densely, are there are different gauge orbits that do this for the
same region. These cannot be separated by open neighbourhoods. As a
result, the quotient space is not Hausdorff. Once again, it should be
emphasized that this is a consequence of the assumption that \emph{both}
the large local translations \emph{and} the large diffeomorphisms are
considered as gauge symmetries. If either of them is dropped, that is,
if one sticks to the original definition of Chern Simons theory, which
does not know anything about large diffeomorphism, or to Einstein
gravity, without singular metrics and without large local translations
as gauge symmetries, then this problem is absent. In the former case,
the states with the particles filling the circle densely are in fact not
close to each other in phase space, and in the second case they are not
gauge equivalent, so in neither case a single gauge orbit fills a region
of phase space densely.

\end{document}